\documentstyle[twoside,fleqn,espcrc2,epsf]{article}
\newcommand{\csw}{c_{\rm sw}}

\def\ca{c_{\rm A}}
\def\cv{c_{\rm V}}

\def\cT{c_{\rm T}}
\def\csw{c_{\rm sw}}

\def\mc{m_{\rm c}}

\def\CF{C_{\rm F}}
\def\cf{\CF}

\def\ba{b_{\rm A}}
\def\bp{b_{\rm P}}
\def\bv{b_{\rm V}}
\def\bs{b_{\rm S}}
\def\bt{b_{\rm T}}

\def\bm{b_{\rm m}}

\newcommand{\AmS}{{\protect\the\textfont2
  A\kern-.1667em\lower.5ex\hbox{M}\kern-.125emS}}

\title{
Computation of the improvement coefficient $\csw$ to 1-loop with
improved gluon actions
\thanks{talk presented by S.~Aoki}}

\author{
S.~Aoki\address{Max-Planck-Institut f\"ur Physik,
F\"ohringer Ring 6, D-80805 M\"unchen, Germany}, 
R.~Frezzotti$^{\rm a}$ and
P.~Weisz$^{\rm a}$
}

\begin{document}

\begin{abstract}
The clover coefficient $\csw$ 
is computed at one loop order of perturbation theory
for improved gluon actions including six-link loops.
The O($a$) improvement coefficients for the dimension three
isovector composite operators bilinear in the quark fields
are also calculated.

\end{abstract}

\maketitle

\setcounter{footnote}{0}

\section{Introduction}

For the on-shell $O(a)$ improvement program with the clover 
action\cite{SW,paperI},
it is important to adjust the clover coefficient
$\csw$, which is currently well-determined only for the Wilson
plaquette gluon action\cite{paperII,paperIII,KarlRainer}. 
Although this is enough for the $O(a)$ improvement,
the so-called renormalization group (RG) improved gluon action\cite{Iwasaki},
combined with the clover action, has been recently employed 
in full QCD simulations
\cite{CppacsF}
to reduce possible $O(a^2)$ errors.
The value of $\csw$ for the RG improved 
action, however, 
was known neither non-perturbatively nor at 1-loop order of 
perturbation theory. Instead a ``perturbative mean field'' value
$\csw = (1-0.8412\beta^{-1} )^{-3/4}$ has been used up to now 
in actual simulations. Therefore it seems desirable to
fully determine $\csw$ at 1-loop level, in order
to be able to estimate how large the errors in this
approximation are.

We sketch here the main points of our computation \cite{afw}
of $\csw$ at 1-loop
order of perturbation theory for gluon actions including six-link loops.
We have also computed the mixing coefficients $c_{\rm X}$
as well as the coefficients $b_{\rm Y}$ of the mass dependent
corrections (needed for $O(a)$ improvement) 
for quark bilinear operators of dimension 3.
We follow the method of refs.\cite{paperII,StefanPeter} to compute these 
quantities and use their notations without further notices. 
More details
can be found in ref.~\cite{afw}.

\section{Gauge actions}
\label{sec:definition}
\begin{figure}[htb]
\vspace{0mm}
\begin{center}
\leavevmode
\epsfxsize=4.0cm
\epsfbox{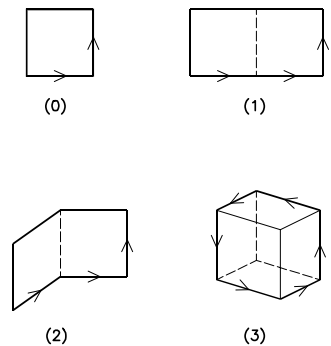}
\end{center}
\vspace{-15mm}
\caption{Elementary loops in ${\cal S}_i,\,\,i=0,1,2,3$}
\label{fig:1loop}  
\vspace{-10pt}
\end{figure}

In this section we specify the gauge actions, focusing
on differences with respect to previous calculations\cite{paperII,StefanPeter}.

The considered gauge action $S[U]$ belongs to a general 
class containing loops up to length 6
\begin{equation}
S[U]={2\over g_0^2}\sum_{i=0}^3 c_i
\sum_{{\cal C}\in {\cal S}_i} {\cal L}({\cal C})
\end{equation}
where the ${\cal S}_i$ denote sets of elementary loops ${\cal C}$
on the lattice as given in fig.~\ref{fig:1loop}, and
${\cal L}({\cal C})={\rm Re Tr}[I-U({\cal C})]$
with $U({\cal C})$ being the ordered product of the link
variables $U_{\mu}(x)$ along ${\cal C}$. The coefficients
$c_i$ are normalized
such that $ c_0+8c_1+16c_2+8c_3=1 $.

\begin{table*}[hbt]
\setlength{\tabcolsep}{0.3pc}
\newlength{\digitwidth} \settowidth{\digitwidth}{\rm 0}
\catcode`?=\active \def?{\kern\digitwidth}
\caption{Improvement coefficients at 1-loop for various gluon actions.
The results for the standard plaquette action are taken from
refs.{\protect
\cite{paperII,StefanPeter}}. 
Tadpole contributions for $\csw^{(1)}$ are also listed. 
}
\label{tab:result}
\begin{tabular*}{\textwidth}{@{}c@{\extracolsep{\fill}}|lllll}
\hline
gauge action& Plaq. & LW & RG1 & RG2 & RG3  \\
\hline
$c_1$ &0.0 &-1/12 &-0.331 & -0.27 &-0.252 \\
$c_3$ &0.0 &\phantom{-}0.0    &\phantom{-}0.0   & -0.04 &-0.17  \\
\hline
$\csw^{(1)}$       & 0.267(1) & 0.196(6) & 0.113(3) & 0.119(5) & 0.109(5)
\\
$\csw^{tad}$ & 0.25     & 0.183 & 0.105 & 0.110 & 0.096 \\
\hline
&\multicolumn{5}{c}{$\times \cf$}   \\
$\ca^{(1)}$ &-0.005680(2)&-0.004525(25) &-0.002846(11) &-0.003017(12) &
-0.002805(20)\\
$\cv^{(1)}$ &-0.01225(1)
&-0.0103(3)&-0.00730(20)&-0.00757(26)&-0.00709(20)\\
$\cT^{(1)}$ &\phantom{-}0.00896(1) &\phantom{-}0.00743(7)
&\phantom{-}0.00505(10) &\phantom{-}0.00526(15) &\phantom{-}0.00496(12) \\
$\bm^{(1)}$ &-0.07217(2)&-0.0576(11)&-0.0382(8)&-0.0395(15) &-0.0353(12)
\\
$\ba^{(1)}$
&\phantom{-}0.11414(4)&\phantom{-}0.0881(13)&\phantom{-}0.0550(4)
&\phantom{-}0.0572(6) &\phantom{-}0.0510(5)  \\
$\bv^{(1)}$
&\phantom{-}0.11492(4)&\phantom{-}0.0884(26)&\phantom{-}0.0551(19)
&\phantom{-}0.0574(19)&\phantom{-}0.0510(21) \\ 
$\bp^{(1)}$
&\phantom{-}0.11484(2)&\phantom{-}0.0889(14)&\phantom{-}0.0558(9)
&\phantom{-}0.0584(10)&\phantom{-}0.0528(8) \\
$\bs^{(1)}$
&\phantom{-}0.14434(4)&\phantom{-}0.1152(22)&\phantom{-}0.0764(16)
&\phantom{-}0.0790(30)&\phantom{-}0.0706(24)\\
$\bt^{(1)}$
&\phantom{-}0.10434(4)&\phantom{-}0.0790(25)&\phantom{-}0.0477(12)
&\phantom{-}0.0502(19)&\phantom{-}0.0444(15) \\ 
\hline
\end{tabular*}
\end{table*}

In our computations we have only considered actions
with $c_2=0$. Apart from the Wilson action ($c_1=c_3=0$), we have 
studied 4 actions, which are specified in table~\ref{tab:result}.

To compute improvement coefficients we work in the framework of
the Schr\"odinger functional (SF), where the theory is defined on hypercubic 
lattices of volume $L^3\times T$ with cylindrical geometry, i.e.
periodic-type boundary conditions in the spatial directions and 
Dirichlet boundary conditions in the time direction.

To ensure $O(a)$ improvement for all on--shell quantities 
in the SF framework, one needs in the gauge action a careful choice 
of weights for
the 
loops that are located near the time boundaries.
Following ref.~\cite{afw}, we rewrite the gauge action as
\begin{equation}
S[U]={2\over g_0^2}\sum_{i=0}^3
\sum_{{\cal C}\in {\cal S}_i}
W_i({\cal C}){\cal L}({\cal C}) \; ,
\end{equation}
where the elements of the classes ${\cal S}_i$ consist of all
loops of the given shape which can be drawn on the cylindrical lattice.
In particular rectangles protruding out of the cylinder
are not included and hence we do not have to specify further
boundary conditions for link variables outside the cylinder\cite{Klassen}.
The weights $W_i({\cal C})$ are set to $c_i$, for $i=0,1,2,3$ and
for all loops excepted the plaquettes ($i=0$) and the rectangles ($i=1$)
near the time boundaries.


Since a non-vanishing background field is required for the calculation of 
$\csw^{(1)}$, we adopt the following choice for the remaining weights:
$W_1({\cal C})=\frac32 c_1$ if the rectangle ${\cal C}$ has exactly 2 links
on a time boundary, $W_1({\cal C})= c_1$ if it has only
1 link on a time boundary and 
$W_{1,2,3}({\cal C})= 0$ if it completely lies on one of 
the time boundaries; $W_0({\cal C})= c_0$ if the plaquette ${\cal C}$
touches a time boundary and $W_0({\cal C})= \frac12$ if it completely lies
on a time boundary.
An advantage of this choice is that
the classical background field, induced by the SF boundary conditions,
is analytically known.

Since no background field is necessary 
for the calculation of $c_{\rm X}^{(1)}$ and $b_{\rm Y}^{(1)}$, 
we have used in this case a simpler choice of weights: 
$W_1({\cal C})= c_1$ if the rectangle ${\cal C}$ just touches one 
of the boundaries and $W_{1,2,3}({\cal C})= 0$ if it completely lies
on one of the time boundaries; 
$W_0({\cal C})= c_0+2c_1$ if the plaquette ${\cal C}$
touches a time boundary and $W_0({\cal C})= \frac12$ if it completely lies
on a time boundary. 

\section{Results and Conclusion}
We present in table \ref{tab:result}
a synthesis of all our 1-loop results, including a comparison with the
case of the standard plaquette action.

We first notice that tadpole contributions to the
$\csw^{(1)}$, denoted $\csw^{tad}$ in the table, give about 90\% of 
the complete 1-loop contributions for all actions considered here.
Therefore the value of $\csw$ taken by the CP-PACS collaboration for their
full QCD simulation with RG1 gauge action\cite{CppacsF} is very close to 
the full one-loop value up to order $g_0^4$: 
$\csw^{pert.} = \csw^{CP-PACS}+0.008 g_0^2 +{\rm O}(g_0^4)$ .

Although only 3 choices of RG improved gauge actions are considered,
it seems that RG improved gauge actions generally give 
a $\csw^{(1)}$ which is a factor 2 to 2.5 smaller
than that of the plaquette action,
while for the perturbative improved action (LW)
the reduction factor is only about $1.35$.
This tendency has already been found in the finite 
part of 1-loop renormalization factors for various quantities \cite{ATNU},
and it is also observed  in table \ref{tab:result} for
other improvement coefficients, the $c_{\rm X}^{(1)}$'s
and $b_{\rm Y}^{(1)}$'s.
Recently the coefficients $c_{\rm X}^{(1)}$ and $b_{\rm Y}^{(1)}$ have been 
calculated
by Taniguchi and Ukawa \cite{TaUk} using a completely different method. 
The two sets of results agree well.

\begin{figure}[htb]
\begin{center}
\vspace{0mm}
\leavevmode
\epsfxsize=7.0cm
\epsfbox{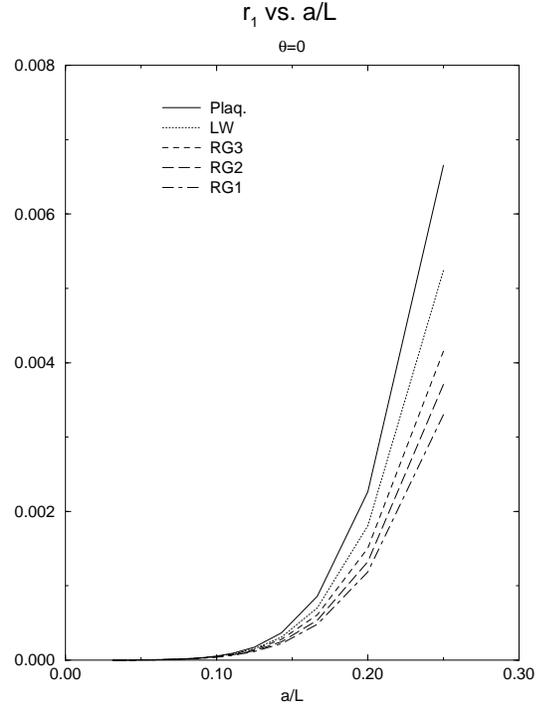}
\end{center}
\vspace{-15mm}
\caption{Remaining cutoff effect at 1-loop in the PCAC mass at
$x_0=T/2$ and $m_0 = \mc$ for various actions.
}
\label{fig:pcac}  
\vspace{-10pt}
\end{figure}

The smallness of 1-loop improvement coefficients for the
RG improved actions does not imply the smallness of the lattice artifacts
for the same actions.
To get some impression of these,
following ref.~\cite{paperII}, 
we consider the 
unrenormalized current quark mass $m$, defined through the PCAC relation
(see eq.(6.13) of ref.~\cite{paperI} ), which is expected to vanish up to 
terms of order $a^2$ at $m_0 = \mc$.
In the perturbative expansion of $m$, 
$ am  = r_0 + r_1 \cf g_0^2 + {\rm O}(g_0^4) \; $,
the value of $r_1$ at $m_0=\mc$ and $x_0=T/2$
represents the magnitude of the remaining
cutoff effect for the particular gauge action at 1-loop order,
since the tree-level contribution 
$r_0$ is independent of the pure gauge action.
In fig.~\ref{fig:pcac} we have plotted $r_1$ for various actions
as a function of $a/L$, taking $T= 2 L$, $\theta = 0$ and 
vanishing boundary gauge field (in which case 
the tree-level contribution actually vanishes). 
At large values of $a/L$ it is observed that this
1-loop lattice artifact 
is indeed smaller for the RG improved gauge actions than
for the LW action, for which it is still smaller than 
for the plaquette action.



\begin{thebibliography}{99}

\bibitem{SW}
B. Sheikholeslami and R. Wohlert,
Nucl. Phys. B259 (1985) 572

\bibitem{paperI}
M. L\"uscher, S. Sint, R. Sommer and P. Weisz,
Nucl. Phys. B478 (1996) 365

\bibitem{paperII}
M. L\"uscher and P. Weisz,
Nucl. Phys. B479 (1996) 429

\bibitem{paperIII}
M. L\"uscher, S. Sint, R. Sommer, P. Weisz and U. Wolff,
Nucl. Phys. B491 (1997) 323

\bibitem{KarlRainer}
K. Jansen and R. Sommer,
Nucl. Phys. B (proc. Suppl. 63A-C (1998) 853
CERN preprint, CERN-TH/98-84, hep-lat/9803017.


\bibitem{Iwasaki}
Iwasaki,
Nucl. Phys. B258 (1985) 141;
UTHEP-118 (1983), unpublished.

\bibitem{CppacsF}
S. Aoki et al. ,
Nucl. Phys. B (proc. Suppl.) 63A-C (1998) 221

\bibitem{afw}
S. Aoki, R. Frezzotti and P. Weisz,
MPI preprint, MPI-PhT/98-48, hep-lat/9808007

\bibitem{StefanPeter}
S. Sint and P. Weisz, 
Nucl. Phys. B502 (1997) 251;
Nucl. Phys. B (proc. Suppl.) 63A-C (1998) 856

\bibitem{Klassen}
T. Klassen,
Nucl. Phys. B509 (1998) 391 

\bibitem{ATNU}
S. Aoki, K. Nagai, Y. Taniguchi, and A. Ukawa,
hep-lat/9802034  

\bibitem{TaUk}
Y. Taniguchi and A. Ukawa,
hep-lat/9806015

\end{thebibliography}
\end{document}